\documentclass[preprint,a4paper,final,showpacs,bibnotes,nofootinbib,notitlepage,pra,tightenlines,twocolumn,10pt]{revtex4}
\usepackage{amssymb}

\usepackage{amsmath}

\newtheorem{theorem}{Theorem}
\newtheorem{acknowledgement}[theorem]{Acknowledgement}

\input{tcilatex}

\begin{document}

\title{Decoherence and relaxation of a qubit coupled to an Ohmic bath
directly and via an intermediate harmonic oscillator}
\author{Xian-Ting Liang\thanks{%
Electronic address: xtliang@ustc.edu}}
\affiliation{Department of Physics and Institute of Modern Physics, Ningbo University,
Ningbo, 315211, China}

\begin{abstract}
Using the numerical path integral method we investigate the decoherence and
relaxation of qubits coupled to an Ohmic bath directly and via an
intermediate harmonic oscillator (IHO). Here, we suppose the oscillation
frequencies of the bath modes are higher than the IHO's. When we choose
suitable parameters the qubits in the two models may have almost same
decoherence and relaxation times. However, the decoherence and relaxation
times of the qubit in the qubit-IHO-bath model can be modulated through
changing the coupling coefficients of the qubit-IHO and IHO-bath and the
oscillation frequency of the IHO.
\end{abstract}

\maketitle

\section{Introduction}

Since Shor's algorithm \cite{Shor} for factoring large numbers, the theory
of quantum computation and quantum information has attracted great recent
interest. It is believed that quantum computers may perform some useful
tasks more efficiently than their classical counterparts. Despite the great
promises of performing quantum computations, however, there are still many
practical difficulties to be resolved before quantum computers might become
available in future. One of the difficulties is that the qubit has too short
decoherence time, which is in fact a central impediment for practical qubit
to be taken as the cell of quantum computers. Solid-state qubits are
considered to be promising candidates for realizing building blocks of
quantum computers because of their integrability and flexibility in the
design. However, the quantum logical gates or registers made up with these
qubits have still shorter decoherence and relaxation times. These motivated
a lot of studies on the decoherence, relaxation and manipulation of the
qubits. Many significant results in the field, not only theoretical but also
experimental, have been achieved. Most of the theoretical research is based
on a spin-boson (SB) model which is supposed to be constructed with a spin
(qubit or two-level system) coupled to a bath. The model has many physical
correspondences and has been widely investigated in recent years \cite%
{Weiss,RevModPhys_59_1,AnnPhys_149_374}. In order to investigate a qubit in
its environment, there is another model which is different from the original
spin-boson one. In the model, the qubit couples to the coordinate $X$ of a
harmonic oscillator, which we shall sometimes call the ``intermediate
harmonic oscillator'' (IHO), and which in turn is coupled to a bath. We call
the model the spin-IHO-bath (SIB) model. Recently, this model receives much
interest in the context of quantum computing with condensed matter systems,
especially with superconducting flux qubit devices, see Ref. \cite%
{ChemPhys_296_333} and within. It is also useful to investigate the
measurement of the solid-state qubit \cite{PhysRevB_68_060503(R)} and
magnetic resonance force microscopy \cite{Nature_430_329,PhysRevB_69_115419}.

On the other hand, the SB and SIB models can be used to describe the
electron transfer in chemical and biological molecules. Many theoretical
investigations in this field use the SB model \cite%
{ChemPhys_310_33,JChemPhys_118_179,JPhysChemA_103_9417,JChemPhys_102_4600}.
Based on Leggett \cite{PhysRevB_30_1208}, Garg \emph{et al.} \cite%
{JChemPhys_83_4491} investigated the SIB model, and obtained a map of it to
the SB one, and obtained the spectral density of the effective bath for the
map. They used this model to study the migration of an electron from one
biomolecule to another, or between two localized sites in the same
biomolecule. It has been shown that the coherence is very important not only
to qubits for making quantum computers but also to electrons for
transferring energy in biological systems \cite%
{Nature_446_782,Science_316_1462}. However, if one wants to know the
decoherence and relaxation behaviors of the two-level systems in their
environment, essentially, the dynamics of the systems needs to be known.

If the qubit energy splitting (denoted by $\Delta $ hereinafter) is not
equal to zero, the two models are not exactly solvable. However, they can be
analyzed using adiabatic renormalization in which a systematic weak damping
approximation must be used. They can also be investigated with some
approximation methods based on the perturbative scheme which also asks for
the systems (qubits) weakly coupling to their environment. Many other
methods \cite{PhysRevA_70_062106,ChemPhys_296_315,RepProgPhys_63_669} for
solving the models have been proposed in recent years, most of which are
based on the Born-Markov approximation. However, it has been pointed out
that the use of the approximation is inappropriate at the large tunneling
amplitude and low temperatures. Recently, some different schemes to solve
the SIB model have also been put forward. Gassmann \emph{et al.} \cite%
{PhysRevB_69_115419} obtain an exactly solvable model from the SIB through
dropping an unimportant term, where an approximation similar to the
Born-Oppenheimer one is used. This method is not successful as the
``dropping term'' is not small enough. So it is important to find out some
methods to accurately estimate the dynamics of the qubits in the two models.
Based on the insight into the dynamics we may understand the decoherence and
relaxation better and may bring forward some schemes on how to suppress
them. It is also of much interested to find out the qubit in which model, SB
or SIB, has longer decoherence and relaxation times. These problems are not
obvious.

An excellent method, accurate numerical path integral method based on the
qusiadiabatic propagator path integral (QUAPI) scheme \cite%
{ChemPhysLett_221_482,JMathPhys_36_2430} may be a suitable tool for solving
the two models. To our problems we choose the iterative tensor
multiplication (ITM) algorithm for the numerical scheme. As Makri \cite%
{ChemPhysLett_221_482,JMathPhys_36_2430} addressed that the method is
non-Markovian and it can make the calculations accurate enough even at very
low temperatures, large tunneling amplitude and strong couplings for which
the Markovian approximation is unsuitable. In this paper we shall use the
tool to investigate the dynamics and then the decoherence and relaxation of
the qubit in the SB and SIB models.

\section{Models and dynamics}

The Hamiltonian of the SB model is%
\begin{equation}
H_{SB}=\hbar \left( \frac{\epsilon }{2}\sigma _{z}+\frac{\Delta }{2}\sigma
_{x}\right) +\tsum_{i}\left[ \frac{p_{i}^{2}}{2m_{i}}+\frac{1}{2}m_{i}\omega
_{i}^{2}\left( x_{i}+\frac{c_{i}\sigma _{z}}{m_{i}\omega _{i}^{2}}\right)
^{2}\right] .  \label{e1}
\end{equation}%
Suppose the bath has an Ohmic spectral density%
\begin{equation}
J_{ohm}\left( \omega \right) =\frac{\pi }{2}\tsum_{i}\frac{c_{i}^{2}}{%
m_{i}\omega _{i}}\delta \left( \omega -\omega _{i}\right) =\frac{\pi }{2}%
\hbar \xi \omega e^{-\omega /\omega _{c}}.  \label{e2}
\end{equation}%
Here, $\xi $ is the dimensionless Kondo parameter\ \cite%
{JChemPhys_102_4600,JChemPhys_102_4611} (the relationship of $\xi $ with the
friction coefficient $\eta $ is $\xi =2\eta /\pi \hbar $ \cite%
{PhysRevA_70_042101}), $\sigma _{i}$ $\left( i=x,\text{ }z\right) $ are the
Pauli matrix, and $\omega _{c}$ is the high-frequency cutoff of the bath
modes. This is a well-known quantum dissipation model and it has been widely
investigated \cite{Weiss,RevModPhys_59_1}.

If we consider the qubit coupling to the coordinate $X$ of a single IHO
which in turn is coupled to a bath, and if we let the couplings be linear,
the Hamiltonian of the SIB system reads%
\begin{eqnarray}
H_{SIB} &=&\hbar \left( \frac{\epsilon }{2}\sigma _{z}+\frac{\Delta }{2}%
\sigma _{x}\right) +\frac{P^{2}}{2M}+\frac{1}{2}M\Omega _{0}^{2}\left(
X+\lambda \sigma _{z}\right) ^{2}  \notag \\
&&+\tsum_{i}\left[ \frac{p_{i}^{2}}{2m_{i}}+\frac{1}{2}m_{i}\omega
_{i}^{2}\left( x_{i}+\frac{\kappa c_{i}X}{m_{i}\omega _{i}^{2}}\right) ^{2}%
\right] ,  \label{e3}
\end{eqnarray}%
where $M$ and $P$ are the mass and momentum of the IHO, and the displacement 
$\lambda $ characterizes the coupling of the qubit to the IHO, and $\kappa
c_{i}$ are the coupling coefficients of the IHO to the bath modes. It is
shown that the system has a one to one map to the following system \cite%
{JChemPhys_83_4491}%
\begin{equation}
H_{SIB}=\hbar \left( \frac{\epsilon }{2}\sigma _{z}+\frac{\Delta }{2}\sigma
_{x}\right) +\tsum_{i}\left[ \frac{\tilde{p}_{i}^{2}}{2\tilde{m}_{i}}+\frac{1%
}{2}\tilde{m}_{i}\tilde{\omega}_{i}^{2}\left( \tilde{x}_{i}+\frac{\tilde{c}%
_{i}\sigma _{z}}{\tilde{m}_{i}\tilde{\omega}_{i}^{2}}\right) ^{2}\right] ,
\label{e4}
\end{equation}%
with an effective spectral density (see Appendix)%
\begin{eqnarray}
J_{eff}\left( \omega \right)  &=&\frac{\pi }{2}\tsum_{i}\frac{\tilde{c}%
_{i}^{2}}{\tilde{m}_{i}\tilde{\omega}_{i}}\delta \left( \omega -\tilde{\omega%
}_{i}\right)   \notag \\
&=&\frac{\pi }{2}\lambda ^{2}\kappa ^{2}\xi \hbar \omega \frac{\Omega
_{0}^{4}}{\left( \omega ^{2}-\Omega _{0}^{2}\right) ^{2}+4\Gamma ^{2}\omega
^{2}},  \label{e5}
\end{eqnarray}%
where $\Gamma =\kappa ^{2}\eta /2M.$ When the bath modes have lower
frequencies than $\Omega _{0}$, the dynamics of the qubit in the bath with
effective spectral density $J_{eff}\left( \omega \right) $ is similar to the
dynamics of the qubit in Ohmic bath with spectral density $J_{ohm}\left(
\omega \right) $, which is widely investigated in recent years \cite%
{AnnPhys_293_15,PhysRevLett_93_267005,JModOpt_47_2905,EurPhysJB_45_405,PhysRevA_65_012309,PhysRevB_65_144516,PhysRevB_72_235321}%
. In this paper we investigate another limited case, namely, the bath modes
have higher frequencies than the oscillation frequency $\Omega _{0}$ of IHO.
The length of the memory times of the baths can be estimated by the
following bath response function%
\begin{equation}
\alpha \left( t\right) =\frac{1}{\pi }\int_{0}^{\infty }d\omega J\left(
\omega \right) \left[ \coth \left( \frac{\beta \hbar \omega }{2}\right) \cos
\omega t-i\sin \omega t\right] .  \label{e6}
\end{equation}%
Here, $\beta =1/k_{B}T$ where $k_{B}$ is the Boltzmann constant, and $T$ is
the temperature. It is shown that when the real and imaginary parts behave
as the delta function $\delta \left( t\right) $ and its derivative $\delta
^{\prime }\left( t\right) ,$ the dynamics of the reduced density matrix is
Markovian. However, if the real and imaginary parts are broader than the
delta function, the dynamics is non-Markovian. The broader the Re$[\alpha
\left( t\right) ]$ and Im$[\alpha \left( t\right) ]$ are, the longer the
memory time will be. The broader the Re$[\alpha \left( t\right) ]$ and Im$%
[\alpha \left( t\right) ]$ are, the more serious the practical dynamics will
be distorted by the Markovian approximation. The memory time of the
effective bath is determined by $\Gamma .$ The larger the $\Gamma $ is, the
shorter the memory time of the effective bath will be. If the memory time of
the bath is short enough, the reduced density matrix may be obtained in
virtue of the Markovian approximation. However, if the memory time of the
bath is too long, the ITM method is in fact inappropriate, saying nothing of
other methods based on Markovian approximation. Clearly, the value of the $%
\Gamma $ may vary according to the difference of the physical systems. For
example, when the persistent-current qubit is measured by a dc SQUID, the
system can be modeled by Eq. (\ref{e4}) with Eq. (\ref{e5}), here $\Gamma
=1/R_{s}C_{s}.$ Typically, $R_{s}=100$ $\Omega ,$ $C_{s}=5\ $pF, so $\Gamma
\sim 10^{11}$, see Ref. \cite{PhysRevB_65_144516}. In this paper we set $%
\Gamma =2.6\times 10^{11}.$ This value of $\Gamma $ makes the calculation
with ITM suitable but other methods based on Markovian approximation
unsuitable.

In Fig. 1 we plot the Re$[\alpha \left( t\right) ]$ and Im$[\alpha \left(
t\right) ]$ as $J\left( \omega \right) =J_{ohm}\left( \omega \right) $, and $%
J_{eff}\left( \omega \right) $, where we set $\lambda \kappa =1050,$ $\xi
=0.01,$ $\Omega _{0}=10\Delta $, $T=0.01$, $\Gamma =2.6\times 10^{11},$ $%
\Delta =5\times 10^{9}$ Hz, the lower-frequency and high-frequency cut-off
of the baths modes $\omega _{0}=11\Delta $, and $\omega _{c}=100\Delta $. It
is shown that the memory time of the Ohmic bath is similar to the one of the
effective bath, namely, $\omega _{c}\tau _{SB}^{m}\sim \omega _{c}\tau
_{SIB}^{m}$. Both baths have shorter effective memory times. This will be
verified in Fig. 2 below.

The dynamics of the qubit is characterized by the time evolution of the
reduced density matrix, obtained after tracing out the bath degrees of
freedom, i.e.,%
\begin{equation}
\rho \left( s^{\prime \prime },s^{\prime };t\right) =\text{Tr}%
_{bath}\left\langle s^{\prime \prime }\right| e^{-i\mathcal{H}t/\hbar
}R\left( 0\right) e^{i\mathcal{H}t/\hbar }\left| s^{\prime }\right\rangle .
\label{e7}
\end{equation}%
In actual cases the initial state of the total system must be entangled even
at the beginning of the evolution. However, at the beginning the
entanglement is very weak, otherwise the qubit must have lost its coherence
and relaxed to its thermal equilibrium in a very short time. This is not our
case investigated in this paper. So, for simplicity of the presentation we
assume that the entanglement (through interaction) between the qubit and its
environment is switched on at $t=0$, i.e., the initial density matrix has
the form \cite%
{ChemPhysLett_221_482,JMathPhys_36_2430,JChemPhys_102_4600,JChemPhys_102_4611,PhysRevA_70_042101}%
\begin{equation}
R\left( 0\right) =\rho \left( 0\right) \otimes \rho _{bath}\left( 0\right) ,
\label{e8}
\end{equation}%
where $\rho \left( 0\right) $ and $\rho _{bath}\left( 0\right) $ are the
initial states of the qubit and bath. The scheme which we use to calculate
the reduced density matrix $\rho (t)$ is a well established ITM algorithm
derived from the QUAPI. It is a numerically exact algorithm and is
successfully tested and adopted in various problems of open quantum systems %
\cite{JChemPhys_102_4600,JChemPhys_102_4611,PhysRevA_70_042101}. For details
of the scheme, we refer to previous works \cite%
{ChemPhysLett_221_482,JMathPhys_36_2430}. The QUAPI asks for the system
Hamiltonian splitting into two parts $H_{0}$ and $H_{env}.$ Here, we take $%
H_{0}=\hbar \left( \frac{\epsilon }{2}\sigma _{z}+\frac{\Delta }{2}\sigma
_{x}\right) $ and $H_{env}=H_{SB}-H_{0}$, or $H_{env}=H_{SIB}-H_{0}$. In
order to make the calculations converge we use the time step $\omega
_{c}\Delta t=0.6,$ which is smaller than the characteristic times of the
qubits in the systems.

\section{Decoherence and relaxation}

The decoherence is in general produced due to the interaction of the quantum
system with other systems with a large number of degrees of freedom, for
example the devices of the measurement or environment. To measure the
decoherence one may use the entropy, the first entropy, and many other
measures, such as the maximal deviation norm, etc. (see for example Refs. %
\cite{JStatPhys_110_957,PhysRevA_69_032311,PhysLettA_328_87}). However,
essentially, the decoherence of an open quantum system is reflected through
the decays of the off-diagonal coherent terms of its reduced density matrix %
\cite{PhysRevB_72_245328}. The decoherence time denoted by $\tau _{2}$
measures the time of the initial coherent terms to their $1/e$ times,
namely, $\rho _{i}\left( n,m\right) \overset{\tau _{2}}{\rightarrow }\rho
_{f}\left( n,m\right) =\rho _{i}\left( n,m\right) /e.$ Here, $n\neq m$, and $%
n,$ $m=0$ or $1$ for qubits. In the following, we investigate the
decoherence via directly describing the evolutions of the off-diagonal
coherent terms, instead of using any measure of the decoherence. Similar to
the decoherence, the relaxation of the qubit can also be investigated with
the diagonal elements of the reduced density matrix. The relaxation time is
denoted by $\tau _{1},$ which measures the time of an initial state to the
final thermal equilibrium state through estimating the diagonal terms of the
reduced density matrix, namely, $\rho _{i}\left( n,n\right) \overset{\tau
_{1}}{\rightarrow }e^{-E_{n}/k_{b}T}$. In the following calculations we
assume the initial state of the environment $\rho _{bath}\left( 0\right)
=\prod\nolimits_{k}e^{-\beta M_{k}}/$Tr$_{k}\left( e^{-\beta M_{k}}\right) .$
As calculating the off-diagonal element $\rho _{12}$ we let $\epsilon
=10\Delta $ which can make the $\rho _{12}$ decay stably. If $\epsilon
\rightarrow \Delta $ the $\rho _{12}$ will decay with some oscillations,
which may affect our judgement on decoherence times. The closer the two
parameters are, the more strongly the matrix elements will oscillate. When
we calculate $\rho _{11}$ we choose parameters $\epsilon =\Delta $ because
the oscillations of the $\rho _{11}$ do not affect our judgement on
relaxation times from the figures. The reader should note that, the increase
of the $\Delta \ $and $\epsilon $ will shorten the decoherence and
relaxation times, so the decoherence time $\tau _{2}$ and the relaxation
time $\tau _{1}$ in the figures are not comparable because they are plotted
in different two sets of the parameters. But, it is clear that the
relaxation time $\tau _{1}$ is longer than the decoherence time $\tau _{2}.$

In the ITM scheme, one should at first choose the $k_{\max }$, so that $%
k_{\max }\omega _{c}\Delta t$ must be larger than the effective memory time $%
\omega _{c}\tau _{SB}^{m}\ $and $\omega _{c}\tau _{SIB}^{m}$ of the baths.
In Fig. 2 we plot the reduced density matrix elements $\rho _{11}$ and $\rho
_{12}$ as $k_{\max }=2,$ $3,$ and $4$ for the SB and SIB models. Here, we
set $\lambda \kappa =1050$, the low- and high-frequency cut-off are $\omega
_{0}=11\Delta $, and $\omega _{c}=100\Delta $. From the Fig. 2 we obtain
that as $\lambda \kappa =1050$ the qubit has almost the same decoherence and
relaxation times in SB and SIB models, and the calculations are in fact
convergent as $k_{\max }\geqslant 3$. It is known that the qubit will show
different decoherence and relaxation when it has different initial states.
We can not obtain general results on the decoherence and relaxation from
some special initial states. Essentially, this limitation is inevitable,
because in order to obtain measures\ of the decoherence and relaxation one
should perform the calculations over all possible initial states in general %
\cite{PhysRevB_72_235321,JStatPhys_110_957}. But, we may figure out some
information on the trend of the decoherence and relaxation of the qubit from
some special results. In particular, we are more interested in some special
results because some schemes on quantum information are based on the usage
of some special initial states. In Fig. 3 we plot the decoherence and the
relaxation of the qubit in SB (a) and SIB (b) models in different initial
states. These states are $\left| \xi _{1}\right\rangle =\left[ \sqrt{1/2},%
\sqrt{1/2}\right] ^{T},$ $\left| \xi _{2}\right\rangle =\left[ \sqrt{3/4},%
\sqrt{1/4}\right] ^{T},$ $\left| \xi _{3}\right\rangle =\left[ \sqrt{6/7},%
\sqrt{1/7}\right] ^{T},$ $\left| \xi _{4}\right\rangle =\left[ \sqrt{12/13},%
\sqrt{1/13}\right] ^{T},$ $\left| \xi _{5}\right\rangle =\left[ \sqrt{29/30},%
\sqrt{1/30}\right] ^{T},$ $\left| \xi _{6}\right\rangle =\left[ \sqrt{59/60},%
\sqrt{1/60}\right] ^{T},$ $\left| \xi _{7}\right\rangle =\left[ 1,0\right]
^{T}.$ As $\lambda \kappa >1050$ (or $\lambda \kappa <1050)$ the decoherence
and relaxation times will be shortened (or lengthened), as shown in Fig. 4.
The frequency of the IHO will strongly affect the qubit decoherence and
relaxation. If the $\Omega _{0}$ of the IHO decreases, the decoherence and
relaxation times will be lengthened, and vice versa. Namely, the farther the 
$\Omega _{0}$ departs from the lower-frequency cut-off of the bath, the
longer decoherence and relaxation times the qubit will have, as evidenced in
Fig. 5.

\section{Conclusions}

In this Letter we have investigated the decoherence and relaxation of a
qubit coupled to an Ohmic bath directly and via an IHO. In our
investigations, we fix the tunneling splitting $\Delta $ of the qubit and
assume that the IHO is far off the resonance to the environment modes, i.e.
the oscillation frequency of the IHO is smaller than the lower-frequency
cut-off of the bath modes. In addition, we suppose that the damping
coefficient $\Gamma $ of a bath to the IHO is moderate. If the quantity is
too small, the memory time of the bath may be very long, thus the methods
based on not only the Markovian approximation but also the ITM scheme are
inappropriate for the investigation of the qubit dynamics. On the contrary,
if the $\Gamma $ is too large, and the memory effects of the bath can be
neglected, the dynamics of the qubit can be investigated in many other ways
based on the Markovian approximation. In spite of these limitation our
investigations have wider appeal because the method is suitable for a wider
range of $\Gamma $, and most practical physical systems have the moderate $%
\Gamma $. By use of the accurate numerical path integral scheme, the ITM
based on the QUAPI method we have obtained the evolutions of the reduced
density matrix elements of qubit in the SB and SIB models. We have obtained
that: (1) When $\lambda \kappa =1050$ the qubit in the SIB model has almost
the same decoherence and relaxation times with the qubit in the SB model. If 
$\lambda \kappa >1050$ the damping of a bath to the IHO or (and) the IHO to
the qubit increase, which results in the decrease of the decoherence and
relaxation times, and vice versa. (2) The decoherence and relaxation times
of the qubit in the SIB model increase with the decrease of the oscillation
frequency $\Omega _{0}$ of the IHO, and vice versa. (3) The decoherence and
relaxation times of the qubit in the SB and SIB models will increase with
the decrease of the $\epsilon $ and $\Delta $, which has not been plotted in
the Letter. The longer decoherence and relaxation times are necessary for
not only the qubits for making the quantum computers but also the electrons
for transferring energy in biological systems. In order to make the qubits
or electrons in the SIB model have longer decoherence and relaxation times
we may try to make the $\Omega _{0}$ and $\lambda \kappa $ smaller.

\begin{acknowledgement}
This project was sponsored by National Natural Science Foundation of China
(Grant No. 10675066) and K.C.Wong Magna Foundation in Ningbo University.
\end{acknowledgement}

\section{Appendix}

In the appendix we derive the effective spectral density expressed with Eq. (%
\ref{e5}). As Ref. \cite{JChemPhys_83_4491} pointed out that the map from
Eq. \ref{e3} to Eq. \ref{e4} does not involve the spin (qubit), the same $%
J_{eff}\left( \omega \right) $ will control the dynamics of a continuous
variable $q$ moving in some potential $U(q)$ and coupled to coordinates $X$
and $\{x_{i}\}$ in the same way as the spin. So we can deduce the $%
J_{eff}\left( \omega \right) $ with the Hamiltonian%
\begin{eqnarray}
H &=&\frac{p_{q}^{2}}{2\mu }+U\left( q\right) +\frac{P^{2}}{2M}+\frac{1}{2}%
M\Omega _{0}^{2}\left( X+\lambda q\right) ^{2}  \notag \\
&&+\tsum_{i}\left[ \frac{p_{i}^{2}}{2m_{i}}+\frac{1}{2}m_{i}\omega
_{i}^{2}\left( x_{i}+\frac{\kappa c_{i}X}{m_{i}\omega _{i}^{2}}\right) ^{2}%
\right] ,  \label{A1}
\end{eqnarray}%
where $p_{q}$ is the momentum conjugate to $q$. Defining $U^{\prime }=dU/dt$
and using the dots for the time derivatives, the classical equations of the
motion are%
\begin{eqnarray}
\mu \ddot{q} &=&-U^{\prime }(q)-M\Omega _{0}^{2}(X+\lambda q)\lambda ,
\label{A2a} \\
M\ddot{X} &=&-M\Omega _{0}^{2}(X+\lambda q)-\tsum_{i}\kappa
c_{i}x_{i}-X\tsum_{i}\frac{\kappa ^{2}c_{i}^{2}}{m_{i}\omega _{i}^{2}},
\label{A2b} \\
m_{i}\ddot{x}_{i} &=&-m_{i}\omega _{i}^{2}x_{i}-\kappa c_{i}X.  \label{A2c}
\end{eqnarray}%
Using the Fourier transforms, Eqs. (\ref{A2a}-\ref{A2c}) can be written as%
\begin{eqnarray}
&&\left( -\mu \omega ^{2}+M\Omega _{0}^{2}\lambda ^{2}\right) q\left( \omega
\right) +M\Omega _{0}^{2}X\left( \omega \right) \lambda   \notag \\
&=&-U_{\omega }^{\prime }\left( q\right) ,  \label{A3a} \\
&&\left( -M\omega ^{2}+M\Omega _{0}^{2}+\tsum_{i}\frac{\kappa ^{2}c_{i}^{2}}{%
m_{i}\omega _{i}^{2}}\right) X\left( \omega \right) +\tsum_{i}\kappa
c_{i}x_{i}\left( \omega \right)   \notag \\
&=&-M\Omega _{0}^{2}\lambda q\left( \omega \right) ,  \label{A3b} \\
x_{i}\left( \omega \right)  &=&-\frac{\kappa c_{i}}{m_{i}\left( \omega
_{i}^{2}-\omega ^{2}\right) }X\left( \omega \right) .  \label{A3c}
\end{eqnarray}%
Insetting Eq. (\ref{A3c}) into Eq. (\ref{A3b}), we have%
\begin{eqnarray}
&&\left( -M\omega ^{2}+M\Omega _{0}^{2}-\omega ^{2}\tsum_{i}\frac{\kappa
^{2}c_{i}^{2}}{m_{i}\omega _{i}^{2}\left( \omega _{i}^{2}-\omega ^{2}\right) 
}\right) X(\omega )  \notag \\
&=&-M\Omega _{0}^{2}\lambda q\left( \omega \right) .  \label{A4}
\end{eqnarray}%
Using the notation%
\begin{equation}
L\left( \omega \right) =-\omega ^{2}\left( M+\tsum_{i}\frac{\kappa
^{2}c_{i}^{2}}{m_{i}\omega _{i}^{2}\left( \omega _{i}^{2}-\omega ^{2}\right) 
}\right) ,  \label{A5}
\end{equation}%
and Eq. (\ref{A4}) we have%
\begin{equation}
X(\omega )=\frac{-M\Omega _{0}^{2}\lambda q\left( \omega \right) }{M\Omega
_{0}^{2}+L(\omega )}.  \label{A6}
\end{equation}%
Insetting Eq. (\ref{A6}) into Eq. (\ref{A3a}), we have%
\begin{eqnarray}
&&\left\{ -\mu \omega ^{2}+\frac{M\Omega _{0}^{2}\lambda ^{2}\left[ M\Omega
_{0}^{2}+L(\omega )\right] }{M\Omega _{0}^{2}+L(\omega )}-\frac{\left(
M\Omega _{0}^{2}\right) ^{2}\lambda ^{2}}{M\Omega _{0}^{2}+L(\omega )}%
\right\} q\left( \omega \right)   \notag \\
&=&-U_{\omega }^{\prime }\left( q\right) .  \label{A7}
\end{eqnarray}%
Introducing the function $K(\omega )$, and making 
\begin{equation}
K(\omega )q\left( \omega \right) \equiv \left[ -\mu \omega ^{2}+\frac{%
M\Omega _{0}^{2}\lambda ^{2}L(\omega )}{M\Omega _{0}^{2}+L(\omega )}\right]
q\left( \omega \right) =-U_{\omega }^{\prime }\left( q\right) .  \label{A8}
\end{equation}%
By using Eqs. (\ref{e2}) and (\ref{A4}) we have 
\begin{eqnarray}
L\left( \omega \right)  &=&-\omega ^{2}\left( M+\kappa ^{2}\int_{0}^{\infty }%
\frac{1}{\omega ^{\prime }\left( \omega ^{\prime 2}-\omega ^{2}\right) }%
\right.   \notag \\
&&\times \left. \sum \frac{c_{i}^{2}}{m_{i}\omega _{i}}\delta \left( \omega
^{\prime }-\omega _{i}\right) d\omega ^{\prime }\right)   \notag \\
&=&-\omega ^{2}\left( M+\frac{2\kappa ^{2}}{\pi }\frac{\pi }{2}%
\int_{0}^{\infty }\frac{J_{ohm}}{\omega ^{\prime }\left( \omega ^{\prime
2}-\omega ^{2}\right) }d\omega ^{\prime }\right)   \notag \\
&=&-\omega ^{2}\left( M+\frac{2\kappa ^{2}\eta }{\pi }\int_{0}^{\infty }%
\frac{\exp (-\omega ^{\prime }/\omega _{c})}{\omega ^{\prime }\left( \omega
^{\prime 2}-\omega ^{2}\right) }d\omega ^{\prime }\right)   \notag \\
&=&-M\omega ^{2}+i\eta \kappa ^{2}\omega e^{-\omega /\omega _{c}}.
\label{A9}
\end{eqnarray}%
Taking the cut-off frequency to be infinity, we have%
\begin{equation}
L\left( \omega \right) =-M\omega ^{2}+i\eta \kappa ^{2}\omega .  \label{A10}
\end{equation}%
Substituting this in Eq. (\ref{A8}), from $J_{eff}\left( \omega \right)
=\lim\limits_{\varepsilon \rightarrow 0+}\func{Im}\left[ K(\omega
-i\varepsilon )\right] $ ($\omega $ is real), we have%
\begin{equation}
J_{eff}\left( \omega \right) =\frac{\pi }{2}\lambda ^{2}\kappa ^{2}\xi \hbar
\omega \frac{\Omega _{0}^{4}}{\left( \omega ^{2}-\Omega _{0}^{2}\right)
^{2}+4\Gamma ^{2}\omega ^{2}}.  \label{A11}
\end{equation}%
Here, $\Gamma =\kappa ^{2}\eta /2M$, and $\xi =2\eta /\pi \hbar .$

\section{Figure captions}

Fig. 1 The response functions of the Ohmic bath and effective bath, where $%
\Delta =5\times 10^{9}$ Hz, $\lambda \kappa =1050,$ $\xi =0.01,$ $\Omega
_{0}=10\Delta $, $T=0.01$, $\Gamma =2.6\times 10^{11},$ the lower-frequency
and high-frequency cut-off of the baths modes $\omega _{0}=11\Delta $, and $%
\omega _{c}=100\Delta .$

Fig. 2 The evolutions of reduced density matrix elements $\rho _{12}$ and $%
\rho _{11}$ in SB and SIB models in different values of $K_{\max }$. Here, $%
\epsilon =10\Delta $ for $\rho _{12}$, $\epsilon =\Delta $ for $\rho _{11}$,
the initial state is $\left| \xi _{1}\right\rangle $ and the other
parameters are the same as in Fig. 1.

Fig. 3 The evolutions of reduced density matrix elements of $\rho _{12}$ and 
$\rho _{11}$ in SB (a) and SIB (b) models in different initial states. These
initial states are given in the context. Here, $K_{\max }=3$, and the other
parameters are the same as in Fig. 2.

Fig. 4 The evolutions of reduced density matrix elements of $\rho _{12}$ and 
$\rho _{11}$ in SIB model in different values of $\kappa \lambda $, the
other parameters are the same as in Fig. 2.

Fig. 5 The evolutions of reduced density matrix elements of $\rho _{12}$ and 
$\rho _{11}$ in SIB model in different values of $\Omega _{0}$, the other
parameters are the same as in Fig. 2.

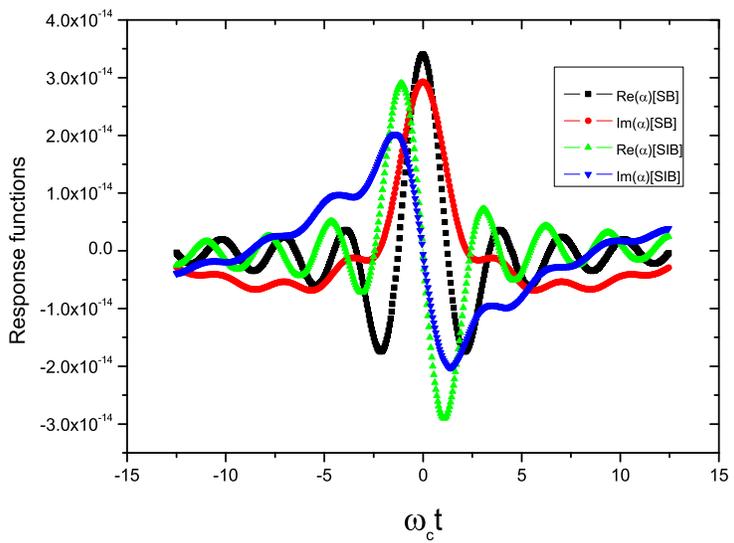
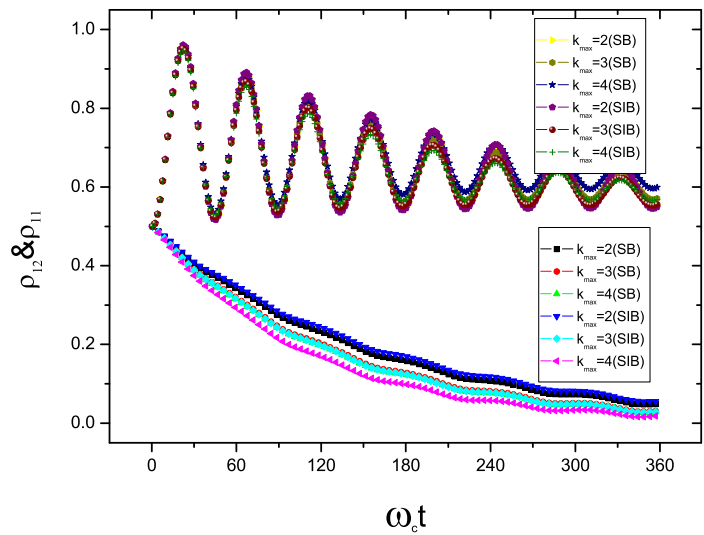
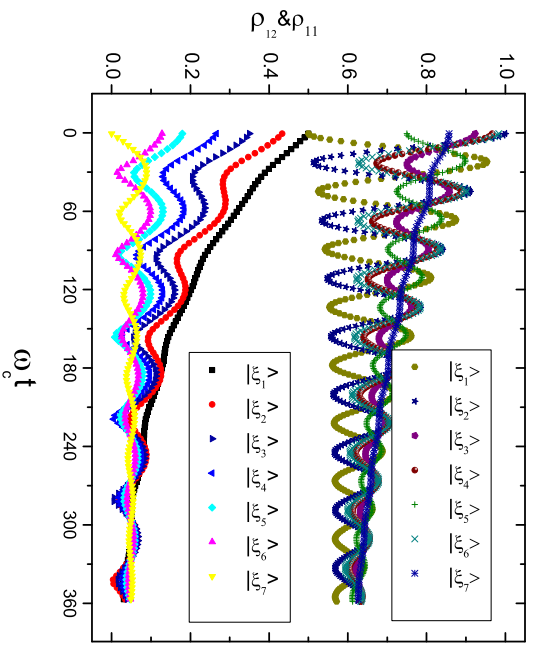
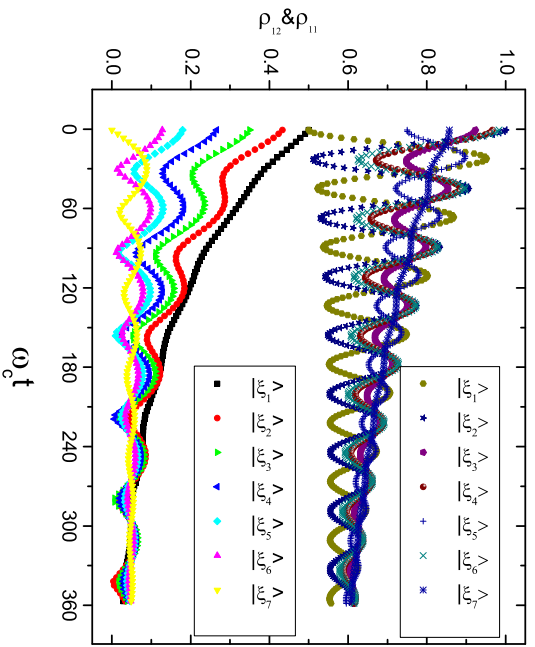
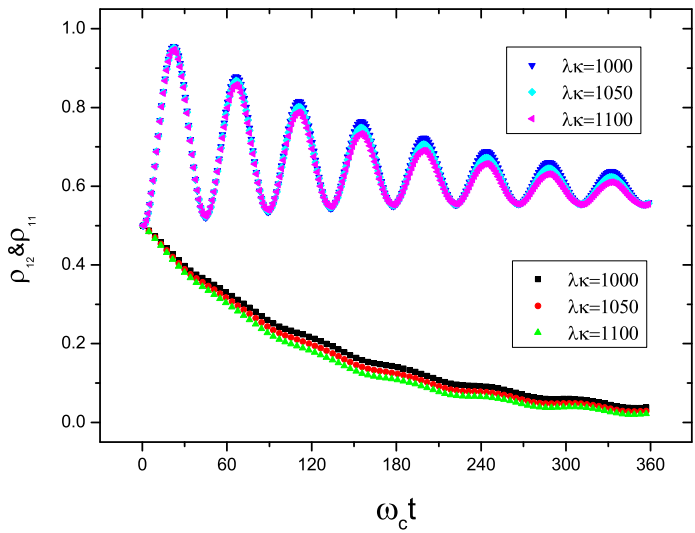
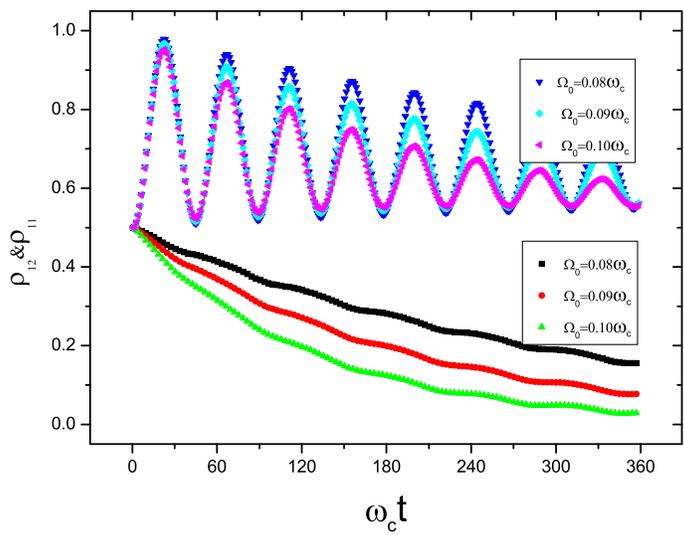
\end{document}